\documentclass[draft,12pt]{article}
\usepackage{amsmath,amssymb,amsthm,bm}
\usepackage{mathbbol,bbm}

\newcommand{\Cx}{{\mathbb C}}

\newcommand{\Rl}{{\mathbb R}}

\newcommand{\idty}{\Eins}
\DeclareMathOperator{\id}{id}

\DeclareMathOperator*{\tr}{Tr}
\newcommand{\<}{\langle}
\renewcommand{\>}{\rangle}
\providecommand{\abs}[1]{\lvert#1\rvert}
\providecommand{\norm}[1]{\lVert#1\rVert}

\renewcommand{\c}[1]{\mathcal{#1}}

\newcommand{\s}[1]{\mathsf{#1}}
\renewcommand{\r}[1]{\mathrm{#1}}

\begin{document}
\begin{center}
{\LARGE Quantum Generalized Subsystems} \\[12pt]
R.~Alicki$^\dagger$, M.~Fannes$^\ddagger$ and M.~Pogorzelska$^\dagger$
\\[6pt]
$^\dagger$ Institute of Theoretical Physics and Astrophysics \\
University of Gda\'nsk, Poland \\[6pt]
$^\ddagger$ Instituut voor Theoretische Fysica \\
K.U.Leuven, Belgium
\end{center}

\medskip\noindent\textbf{Abstract}
We propose a new formalism of quantum subsystems which allows to unify the existing and new methods of reduced description of quantum systems. The main mathematical ingredients are completely positive maps and correlation functions. In this formalism generalized quantum systems can be composed and there is a notion of generalized entanglement. Models of fermionic and bosonic systems and also quantum systems described by the SU(2) symmetry are studied.
\noindent 

\section{Introduction}
\label{s1}

The reduced description of many-particle systems in terms of a relatively small number of parameters is a main tool in Statistical Physics. There exist several schemes of reductions leading to different mathematical structures and involving reduced dynamics as well. The theory of quantum open systems within the operator algebra formalism offers perhaps the richest example of such a description.

Historically, the first general and abstract approach to reduced dynamics of quantum systems was introduced by Nakajima~\cite{na}, Zwanzig~\cite{zw}, and Prigogine~\cite{pr}. It is called the projection technique and is based on a projector operator $\c P$, i.e\ an operator that satisfies $\c P^2 =\c P$. The states of the total system are elements of a Banach space $\c B$ and $\c P$ projects onto a subspace $\c B_0$ that contains the states of the subsystem, i.e.\ the reduced system. The purpose of the projector is to eliminate the irrelevant  freedoms of the so-called environment, reservoir, or heat bath. This leads for the reduced dynamics either to an integro-differential equation with a certain memory kernel or, using the time-convolutionless approach~\cite{st}, to differential equations for the states of the subsystem. 

Another type of reduction can be traced back to the Boltzmann derivation of linear and nonlinear kinetic equations for a gas of particles. Here the complete description in terms of $N$-particle probability distributions is replaced by a density of particles in the single-particle phase space. In the limit of large $N$ and low density one obtains a closed equation of motion. A similar approach in quantum mechanics leads to Hartree type equations for Hamiltonian dynamics and their extensions to linear and nonlinear quantum dynamical semigroups for many-body open systems~\cite{AlMe}.

The aim of this paper is to present a unifying formalism for reduced descriptions in terms of generalized subsystems (GS). In contrast to the projection technique, where only the linear structure of the underlying Banach spaces is retained, we heavily use the algebraic structure of quantum theory. The projection operator $\c P$ is replaced by the dual of a completely positive map $\Phi$ from the algebra of observables of the generalized subsystem, often finite dimensional, to that of the total system. 

The algebraic formalism allows for a rich structure of the GS's including the notion of positivity and the possibility of producing composed GS's with their generalized entanglement. This should be compared with a very different approach introduced in~\cite{v3}. The formalism of GS not only unifies several known instances of reduced descriptions but it provides also new examples like Lie algebraic GS or a quantum like formalism for classical systems.  
 
The paper is organized as follows: generalized subsystems are introduced in Section~\ref{s2}. In Section~\ref{s3} we show how the GS formalism unifies common reduced descriptions such as quantum open systems, coarse-graining, quasi-free boson and fermion systems, and mean-field models. Section~\ref{s4} deals with two less common examples: systems that come with a SU(2) symmetry and a reduced description of the Koopman formalism~\cite{aa}. Finally, composition and entanglement in space and time are briefly introduced in Section~\ref{s5}.

\section{Generalized subsystems}
\label{s2}

We assume in general that the total, usually complex, many-body system is described by a unital C*-algebra $\c A$ its hermitian elements corresponding to bounded observables. A generalized subsystem is determined by a linearly independent family 
\begin{equation}
\s V = \bigl( v_1, v_2, \ldots, v_d \bigr)
\end{equation} 
of elements of $\c A$ called a partition. Introducing the element 
\begin{equation}
m := \sum_i^d v_i^*v_i
\label{gsbas}
\end{equation}
we shall distinguish between partitions of unity where $m = \idty$ and general ones.

A partition $\s V$ generates a completely positive map $\Phi$ from the algebra of complex matrices of dimension $d$, denoted by $\c M_d$ to $\c A$ 
\begin{equation}
\Phi(A) := \sum_{ij} A_{ij}v_i^*v_j,\enskip A = [A_{ij}] \in \c M_d.
\label{lin}
\end{equation}
Note that $\Phi$ is unity preserving if and only if $m = \idty$. The set of states, i.e.\ the linear, positive and normalized functionals on $\c A$, is denoted by $\c S(\c A)$. We consider the pull-back map $\Phi^*$ from $\c S(\c A)$ to the set of positive functionals on $\c M_d$
\begin{equation}
\Phi^*(\omega) = \omega \circ \Phi.
\label{qu:cg:pb} 
\end{equation}
We can identify the functional $\Phi^*(\omega)$ with a positive $d\times d$ matrix $D^\omega$ through
\begin{equation}
\Phi^*(\omega)(A) = \tr \bigl( D^\omega A \bigr),\enskip A \in \c M_d
\end{equation}
and therefore view $\Phi^*$ as an affine map from $\c S(\c A)$ to $\c M_d^+$.  The matrix $D^\omega$ is called a correlation matrix and its entries are easily computed 
\begin{equation}
D^\omega_{ij} = \omega(v_j^*v_i).
\label{} 
\end{equation}
The image of $\Phi^*$ is called a reduced state space and denoted by $\c S(\c A,\s V)$. A reduced state space is easily seen to be a closed convex subset of $\c M_d^+$. In particular, the extreme points of $\c S(\c A,\s V)$ are images of pure states $\omega$ on $\c A$ but generally the converse is not true. The detailed geometrical structure of $\c S(\c A,\s V)$ is one of the problems which should be solved for particular examples.

Two remarks are in order: in some cases the assumption that $\c A$ is a C*-algebra can be lifted and partitions in unbounded elements can be considered. This leads to domain problems, e.g.\ one must choose a suitable Hilbert space representation of the global system and a subset of states $\omega$ for which $D^\omega $ exist. Similarly, with some mathematical care, one can extend the definition of GS beyond finite partitions, allowing for countably or even continuous partitions. 

The map $\Phi^*$ is generally not one to one so that infinitely many states on $\c A$ are mapped on a same $D$, leading to a proper reduction in the description. Still, the states in $\c S(\c A, \s V)$ can sometimes encode to a high degree of approximation the state of the global system. This can be modelled by a embedding map 
\begin{equation}
\Psi: \c S(\c A, \s V) \to \c S(\c A)
\end{equation} 
which is even possibly non-linear. Consistency is expressed by the requirement 
\begin{equation}
\Phi^* \circ \Psi = \id.
\end{equation} 

A reasonable basis for the choice of an embedding is the maximal entropy principle, see~\cite{IKO}: among all states $\omega$ on $\c A$ that return a given $D$, i.e.\ such that $\Phi^*(\omega) = D$, we choose for $\Psi(D)$ the state of maximal entropy. This presupposes both the existence of an entropy $\s S$ on the global algebra and the uniqueness of the constrained maximum. Suppose e.g.\ that 
\begin{equation}
\s S(\omega) = - \tr \omega \log \omega
\end{equation} 
then
\begin{equation}
\Psi(D) = \c Z(\alpha)^{-1} \exp\Bigl\{ -\sum_{ij} \alpha_{ij} v^*_i v_j \Bigr\}
\end{equation}
where
\begin{equation}
\c Z(\alpha) = \tr \exp\Bigl\{ -\sum_{ij} \alpha_{ij} v^*_i v_j \Bigr\}
\end{equation}
and where $\alpha$ is chosen in such a way that
\begin{equation}
D = \Phi^* \Bigl( \c Z(\alpha)^{-1} \exp\Bigl\{ -\sum_{ij} \alpha_{ij} v^*_i v_j \Bigr\} \Bigr).
\end{equation}

An embedding map $\Psi$ allows to construct a reduced dynamics of the GS. E.g.\ if $\bm\alpha = \{\alpha_t \,|\, t\in\Rl\}$ is the Heisenberg evolution of the global system then we can evolve a $D(0) \in \c S(\c A,\s V)$ as
\begin{equation}
D(t) = \Phi^*(\Psi(D(0)) \circ \alpha_t).
\end{equation}
There is no reason to expect a simple dynamical equation for $D(t)$. Nevertheless, simple closed differential equations can be obtained as limiting cases, scaling properly the environment, the map $\Psi$ and the evolution $\bm\alpha$. Well-known examples are the Markovian semigroup evolution obtained through the weak-coupling limit, Hartree-Fock equations, non-linear mean-field equations, \dots

\section{Common generalized subsystems}
\label{s3}

In this section we briefly rephrase some well-known reduced descriptions of quantum systems in terms of GS.

\subsection{Open quantum systems}

The Hilbert space of the total system is the tensor product $\c H_{\r S} \otimes \c H_{\r E}$ of the Hilbert space $\c H_{\r S}$ of the small system of interest and the Hilbert space of the environment $\c H_{\r E}$. The algebra $\c A$ of global observables is that of all bounded operators of the total system, $\c A = \c B\bigl( \c H_{\r S} \otimes \c H_{\r E} \bigr)$. To a given state $\omega$ of the total system a reduced density matrix $D^\omega$ of the subsystem $\r S$ is assigned through the partial trace
\begin{equation}
D^\omega = {\tr}_{\r E}\, \omega. 
\label{rdm}
\end{equation}
Here the state $\omega$ is identified with its corresponding density matrix which is still denoted by the same symbol $\omega$.

This reduced picture can easily be handled in terms of GS by introducing a partition $\bigl(v_1, v_2, \ldots \bigr)$ with
\begin{equation}
v_j = |\varphi\>\<j| \otimes \idty_{\r E}. 
\label{rdm1}
\end{equation}
Here $\{|j\>\}$ is an orthonormal basis in $\c H_{\r S}$ and $\varphi \in \c H_{\r S}$ is an arbitrary normalized vector. The corresponding map $\Phi$ is given by
\begin{equation}
\Phi(A) = \sum_{ij} A_{ij} v_i^*v_j = A \otimes \idty_{\r E}
\end{equation}
and $\Phi^*$ is the partial trace with respect to the environment
\begin{equation}
\Phi^*(\omega)(A) = \omega\bigl( A \otimes \idty_{\r E} \bigr),\enskip A \in \c B\bigl( \c H_{\r S} \bigr). 
\end{equation}
Obviously, for this example the reduced state space $\c S(\c A,\s V)$ consists of all density matrices on $\c H_{\r S}$. It is also clear that $\Phi^*$ is not injective as there are many ways to extend a state on $\c B(\c H_{\r S})$ to the total system.

In order to obtain a well-defined reduced dynamics one starts by extending an arbitrary state $D$ of the system to a global state $\omega = D \otimes \omega_{\r E}$, i.e.\ 
\begin{equation}
\Psi(D(0)) = D(0) \otimes \omega_{\r E}.
\end{equation} 
Here $\omega_{\r E}$ is a suitably chosen reference state of the environment. Obviously $\Phi^* \circ \Psi = \id$. This embedding of states of the system in global states yields a reduced dynamics for  the states of the system
\begin{align}
D(0) \mapsto D(t) 
&= \Phi^* \bigl( U(t) \Psi(D(0)) U(t)^* \bigr) \\
&= {\tr}_{\r E}\, \Bigl(U(t) \bigl( D(0) \otimes \omega_{\r E} \bigr) U^*(t) \Bigr)
\end{align}
where $\{U(t) \,:\, t \in \Rl\}$ is the reversible evolution of the global system. This reduced dynamics is generally very complicated and highly non-Markovian. However, in the regime of weak interaction between system and environment it is governed by a Markovian master equation of standard form~\cite{al} 
\begin{equation}
\frac{d}{dt}D = -i[H \,,\, D] + \frac{1}{2}\, \sum_\alpha \Bigl( [L_\alpha \,,\, D L^*_\alpha] + [L_\alpha D \,,\, L^*_\alpha] \Bigr). 
\label{Lin}
\end{equation}
This scheme has many applications in various fields of physics including quantum information processing in the presence of noise.

\subsection{Coarse graining}

Quite often one is only interested in the occupation probabilities of certain energy levels or groups of almost degenerate energy levels. These are generated by a family of orthogonal projectors $\bigl(P_1, P_2, \ldots \bigr)$  with
\begin{equation}
P_j^* = P_j,\enskip P_iP_j = \delta_{ij} P_i, \enskip\text{and } \sum_j P_j = \idty.
\end{equation}
The corresponding probabilities are
\begin{equation}
p_j = \tr (D P_j)
\end{equation}
where $D$ is the reduced density matrix of the system.

Those probabilities can also be described by a GS defined by a partition $\bigl( P_1 \otimes \idty_{\r E}, P_2 \otimes \idty_{\r E}, \ldots \bigr)$. The correlation matrices are now always diagonal and given by
\begin{equation}
D^\omega_{ij} = p_i \delta_{ij} \enskip\text{with}\enskip p_j = \omega\bigl( P_j \otimes \idty_{\r E} \bigr). 
\label{pauli}
\end{equation}

Again, under certain assumptions and using a Markovian approximation, one can derive the Pauli master equation for the probabilities 
\begin{equation}
\frac{d}{dt}  p_j = \sum_k \bigl( a_{jk}p_k - a_{kj}p_j \bigr).
\label{pauli1}
\end{equation}
Here $a_{jk} \ge 0$ are transition probabilities per unit time typically computed in terms of time-dependent perturbation theory, e.g.\ Fermi's Golden Rule.

\subsection{One particle description for fermions and bosons}

Assume that the system $\r S$ is not small but consists of many particles, fermions or bosons, described annihilation and creation operators associated to a single particle orthonormal basis $\{|k\rangle\}$. They satisfy canonical anticommutation or commutation relations
\begin{equation}
a_k  a^*_\ell \pm a^*_\ell a_k = \delta_{kl} \enskip\text{and}\enskip a_k a_\ell \pm a_\ell a_k = 0.
\label{ccr}
\end{equation}
Now the reduction to the single particle description is possible if only additive observables of the form
\begin{equation}
b := \sum_{\ell,k} B_{k\ell} a^*_k a_\ell,\enskip B = [B_{k\ell}]\enskip\text{Hermitian}
\label{1par}
\end{equation}
are relevant. Therefore, instead of the many-particle density matrix $\rho$ on fermionic or bosonic Fock space  $\c F_\pm$ the 1-particle density matrix $Q = [Q_{k\ell}]$ is used
\begin{equation}
Q_{k\ell} = \tr \bigl(\rho a^*_k a_\ell \bigr) \enskip\text{or}\enskip \tr (Q B) = {\tr}_{\c F_\pm}(\rho\, b).
\label{1par1}
\end{equation}
This 1-particle density matrix $Q$ is positive but normalized to the average number of particles in the system and not to 1. 

This reduction can again be phrased in terms of a correlation matrix choosing the elements of the partition as 
\begin{equation}
v_j = a_j. 
\label{qf}
\end{equation}
I.e.\ 
\begin{equation}
\Phi(B) = \sum_{k\ell} B_{k\ell} a^*_k a_\ell \enskip\text{and}\enskip \Phi^*(\omega) = \bigl[ \omega(a^*_j a_i) \bigr] =: \bigl[ \< i \,,\, Q_\omega j\> \bigr].
\end{equation}
The operator $Q_\omega$ is called a symbol, it is positive semidefinite and satisfies additionally $Q_\omega \le \idty$ for fermions. There is a natural but non-linear map $\Psi$ from the symbol space to the state space of the full system
\begin{equation}
\Psi(Q) = \omega_Q
\end{equation}  
where $\omega_Q$ is either the fermionic or the bosonic quasi-free state with symbol $Q$
\begin{equation}
\omega_Q^\pm \bigl( a^*_{i_1} \cdots a^*_{i_k} a_{j_k} \cdots a_{j_1} \bigr) = {\det}_{\pm} \Bigl( \bigl[ \< j_a \,,\, Q i_b \> \bigr] \Bigr)
\end{equation}
with $\det_-$ equal to the permanent. It is again immediate that $\Phi^* \circ \Psi = \id$. As is the general open quantum system setting, the reduced dynamics is quite complicated.

Quite often, however, a system can be well-modelled by essentially noniteracting quasi-particles and the leading dissipative effects are well-approxima-\\ted by processes of quasi-particle decay and production. In these cases the Markovian master equation~(\ref{Lin}) possesses a particularly simple form~\cite{al}
\begin{equation}
\begin{split}
\frac{d}{dt}\rho = 
&-i[H \,,\,\rho] \\
&+ \frac{1}{2}\, \sum_{k,\ell} \Bigl\{\gamma_{k\ell} \bigl( [a_k \,,\, \rho a^*_\ell] + [a_k\rho \,,\, a^*_\ell] \bigr) + \kappa_{k\ell} \bigl( [a^*_k \,,\, \rho a_\ell] + [a^*_k\rho \,,\, a_\ell] \bigr) \Bigr\}
\end{split}
\label{qfree}
\end{equation}
where $H = \sum_k \epsilon_k a^*_k a_k$ with decay matrix $\gamma = [\gamma_{k\ell}] \ge 0$ and production matrix $\kappa = [\kappa_{k\ell}] \ge 0$. The solution of the master equation~(\ref{qfree}) is a quasi-free dynamical semigroup~\cite{al}. This description can e.g.\ be used to deal with the following situations: \\
a) decay and production of unstable elementary particles, nuclei, quasi-particles, \dots \\
b) propagation of a quantum electromagnetic field in media in a linear regime \\
c) transitions between localized electronic states and a large number of low lying states accompanied by emission and absorption of energy \dots

Inserting~(\ref{qfree}) into~(\ref{1par1}) one can easily derive a closed evolution equation for the 1-particle density matrix $Q$
\begin{equation}
\frac{d}{dt}Q = -i[h \,,\, Q] - \frac{1}{2}\,\{(\gamma \pm \kappa) \,,\, Q\} + \kappa
\label{1pevo}
\end{equation}
where $h := \sum_k \epsilon_k |k\rangle \langle k|$.

Note that for bosons the operators $v_j = a_j$ are unbounded and, moreover, in both the bosonic and fermionic case the set of indices $\{j\}$ may be infinite, see the remarks at the end of Section~\ref{s2}.

\subsection{Mean field models}

The simplest setting is that of systems of $N$ identical but distinguishable particles. The mean field approximation relies on permutation symmetry instead of the usual translation symmetry. Because of this huge symmetry group the set of invariant states becomes quite small. In the limit of large $N$ the set of permutation invariant states, called exchangeable states, reduces to mixtures of permutation invariant product states. Therefore an exchangeable state can be seen as a probability measure on the density matrices of a single particle. This is called de~Finetti's theorem in the classical case and St\o rmer's theorem for quantum systems. The reduction map is defined by
\begin{equation}
\Phi(A) = \frac{1}{N}\, \bigl( A \otimes \idty \cdots \otimes \idty + \cdots + \idty \otimes \cdots \otimes \idty \otimes A \bigr)
\end{equation} 
where $A$ is a one-particle observable. It is easily seen that the reduced state space is just the full set of single particle states and that the maximal entropy embedding is given by
\begin{equation}
\Psi(D) = D \otimes D \otimes \cdots \otimes D.
\end{equation} 

The dynamics generated by an $N$-particle Hamiltonian of the form
\begin{equation}
H_N = \sum_{i=1}^N h^{(1)}_i + \frac{1}{2N}\, \sum_{i \ne j} h^{(2)}_{ij}
\end{equation}
will preserve the permutation symmetry of initial states. Here $h^{(1)}$ and $h^{(2)}$ are Hermitian one and two particle interactions. It then follows that the reduced dynamics is described by a non-linear evolution equation in the limit of large $N$
\begin{equation}
\frac{d\ }{dt} D = -i \bigl[ h^{(1)} \,,\, D \bigr] -i \bigl[ {\tr}_2 h^{(2)} \,,\, D \bigr]
\end{equation}
where ${\tr}_2$ denotes the partial trace over the second factor in $\c H \otimes \c H$, $\c H$ being the one-particle space. General Markovian dynamics can be handled in a similar way.

\section{More examples of generalized subsystems}
\label{s4}

We discuss now two classes of GS which go beyond the standard schemes of reduced description:  Markovian open quantum systems that come with a representation of a Lie algebra and a quantum like picture of classical systems.

\subsection{Lie algebraic open systems}

Consider a quantum open system containing a Lie algebra $\c A_L$ of operators spanned by basis elements $X_m = X_m^*$ satisfying the commutation relations
\begin{equation}
[X_m \,,\, X_n] = \sum_k c_{mnk} X_k.
\label{lie}
\end{equation}
The operators $X_j$ define a partition $\s V = \{v_j = X_j\otimes\idty_E\}$. 

Assume that the dynamics of the open system is governed by a standard Markovian master equation~(\ref{Lin}) with $L_\alpha = L^*_\alpha$. If both the Hamiltonian $H$ and the operators $L_\alpha$ belong to $\c A_L$ then the Heisenberg equation of motion for a product $X_m X_n$ reads
\begin{align}
\frac{d\ }{dt} (X_m X_n) 
&= i X_m [H \,,\, X_n] +  i [X_m \,,\, H] X_n -\frac{1}{2} \sum_\alpha \Bigl(
[L_\alpha \,,\, [L_\alpha \,,\, X_m]] X_n 
\nonumber \\
&\phantom{=\ }+ X_m [L_\alpha \,,\, [L_\alpha \,,\, X_n]] + 2 [L_\alpha \,,\, X_m] [L_\alpha \,,\, X_n] \Bigr) \\ 
&= \sum_{k\ell} a(mn;k\ell) X_k X_\ell.
\label{liedyn}
\end{align}
This yields a closed evolution equation for the correlation matrix 
\begin{equation}
\frac{d\ }{dt} D_{nm} = \sum_{k\ell} a(mn;k\ell) D_{\ell k}.
\label{lieev}
\end{equation}

\subsection{Angular momentum spaces}

In the following we consider quantum systems with irreducible representations of $su(2)$ Lie algebra given by angular momentum operators. We shall denote by $\bm{J^{(\ell)}} = \bigl( J^{(\ell)}_1, J^{(\ell)}_2, J^{(\ell)}_3 \bigr)$ the irreducible spin $\ell$ representation of the three standard generators of the rotation group where $\ell$ takes values in $\{0, \frac{1}{2}, 1, \ldots\}$. The $J^{(\ell)}_i$ are matrices of dimension $2\ell+1$ which satisfy the relations
\begin{equation}
\bigl( J^{(\ell)}_\alpha \bigr)^* = J^{(\ell)}_\alpha,\enskip \bigl[ J^{(\ell)}_\alpha\,,\, J^{(\ell)}_\beta \bigr] = i\, {\epsilon_{\alpha\beta}}^\gamma J^{(\ell)}_\gamma,\enskip \text{and } \bm{J^{(\ell)}} \cdot \bm{J^{(\ell)}} = \ell(\ell+1) \idty. 
\label{ang}
\end{equation}
Here, ${\epsilon_{\alpha\beta}}^\gamma$ is the totally antisymmetric tensor with ${\epsilon_{12}}^3 = 1$. The $J^{(\ell)}_\alpha$ are for a given $\ell$ up to a unitary transformation uniquely determined by the relations~(\ref{ang}) and we shall use the standard convention that $J^{(\ell)}_3$ is diagonal in the standard basis of $\Cx^{2\ell+1}$. To obtain operational partitions of unity we have to renormalize the generators
\begin{equation}
\bm{j^{(\ell)}} = \frac{1}{\sqrt{\ell(\ell+1)}}\, \bm{J^{(\ell)}}. 
\label{ang2}
\end{equation} 
A reduced state description in terms of $\bm{j^{(\ell)}}$ can be practically useful for $\ell \gg 1$, however, the complete description of its structure is here only illustrated for the simplest cases $\ell = 1/2, 1, \infty$.

\subsubsection{Spin 1/2}

The normalized spin 1/2 generators have the form
\begin{equation}
j_1 = \frac{1}{\sqrt 3}\, \begin{bmatrix} 0 & 1 \\1 & 0 \end{bmatrix},\enskip j_2 = \frac{1}{\sqrt 3}\, \begin{bmatrix} 0 & -i\\ i & 0 \end{bmatrix}, \enskip \text{and }j_3 = \frac{1}{\sqrt 3}\, \begin{bmatrix} 1 & 0 \\ 0 & - 1 \end{bmatrix}.
\label{spin1/2}
\end{equation}
The corresponding completely positive map $\Phi$, see~(\ref{lin}), sends a matrix $A$ of dimension 3 into a matrix of dimension 2
\begin{equation}
\Phi(A) = \sum_{\alpha\beta=1}^3 A_{\alpha\beta}\, j_\alpha\, j_\beta.
\label{cg:spin1/2}
\end{equation}

The following characterization of positive semi-definite matrices will be useful. Let $A$ be a square matrix of dimension $d$ and denote by $(\lambda_1, \lambda_2, \ldots, \lambda_d )$ its eigenvalue list repeated according to algebraic multiplicities. The elementary symmetric invariant of order $k$ is given by
\begin{equation}
e_k = \sum_{\substack{\Lambda \subset \{1,2,\ldots,d\} \\  \#(\Lambda) = k}}\enskip \prod_{j\in\Lambda} \lambda_j.
\label{esi}
\end{equation} 
Then $A$ is positive semi-definite if and only if $A = A^*$ and all $e_k(A) \geq 0$ for $k = 1,2,\ldots,d$.\\

The reduced state space consists of all matrices D that satisfy the following condition
\begin{equation}
\tr D = 1 \enskip\text{and}\enskip \tr (DA) \ge 0 \text{ whenever } \Phi(A) \ge 0. 
\label{cond}
\end{equation} 
Note that a $D$ which satisfies~(\ref{cond}) is automatically a density matrix, indeed $\Phi(A) \ge 0$ whenever $A \ge 0$ because $\Phi$ is CP.

One can verify, using the characterization of positivity given in~(esi), that $\Phi(A) \in \c M_2$ is positive if and only if $A \in \c M_3$ is of the form
\begin{equation}
A = \begin{bmatrix} x_1 &a_3 + i \lambda_{12} &a_2 + i \lambda_{13} \\ a_3 + i \lambda_{21} &x_2 & a_1 + i \lambda_{23} \\ a_2 + i \lambda_{31} & a_1 + i \lambda_{32} & x_3 \end{bmatrix}
\label{c1}
\end{equation}
with
\begin{align*}
&a_i \in \Rl,\enskip x_i \in \Cx \enskip\text{and } x_1 + x_2 + x_3 \ge 0, 
\nonumber \\
&\lambda_{ij} \in \Rl \enskip\text{and } (x_1 + x_2 + x_3)^2 \geq (\lambda_{12}-\lambda_{21})^2 + (\lambda_{13}-\lambda_{31})^2 + (\lambda_{23}-\lambda_{32})^2.
\end{align*}

Imposing that $\tr(AD) \ge 0$ for all such $A$ implies that 
\begin{equation}
D = \begin{bmatrix} \frac{1}{3} &i\alpha_3 &-i\alpha_2 \\-i\alpha_3 &\frac{1}{3} &i\alpha_1 \\i\alpha_2 &-i\alpha_1 &\frac{1}{3} \end{bmatrix}
\end{equation} 
where 
$\bm\alpha := (\alpha_1,\alpha_2,\alpha_3) \in \Rl^3$ satisfies $\norm{\bm \alpha} \le \frac{1}{3}$. So the reduced state space in this example is a ball in $\Rl^3$ and hence affinely isomorphic to the state space of a qubit (the Bloch ball). The extreme points are parametrized by $\norm{\bm\alpha} = 1/3$, the corresponding density matrices have eigenvalues $(2/3,1/3,0)$ and are therefore not pure.

The explicit form of $\Phi^*$ is
\begin{equation}
\Phi^*(\rho) = \frac{1}{3}\, \begin{bmatrix} 1 &-ix_3 &ix_1 \\ix_3 &1 &ix_2 \\-ix_1 &ix_2 &1 \end{bmatrix},
\end{equation}
where the qubit density matrix $\rho$ is expanded in Bloch notation
\begin{equation}
\rho = \frac{1}{2}\, (\idty + \bm x \cdot \bm \sigma),\enskip \bm x = (x_1,x_2,x_3) \in \Rl^3 \text{ and } \norm{\bm x} \le 1. 
\end{equation} 

As in this example the dimension of the reduced system is actually larger than that of the original there cannot exist a map $\Psi$ from the reduced system to the full system which is a right inverse of $\Phi^*$. 

\subsubsection{Spin 1}

The normalized generators of the spin 1 representation define the operational partition of unity $\bigl( j_1,j_2,j_3 \bigr)$ with
\begin{equation}
j_1 = \frac{1}{2}\, \begin{bmatrix}0 &1 &0 \\1 &0 &1 \\0 &1 &0\end{bmatrix},\enskip
j_2 = \frac{1}{2}\, \begin{bmatrix}0 &-i &0 \\i &0 &-i \\0 &i &0\end{bmatrix},\enskip \text{and }
j_3 = \frac{1}{\sqrt2}\, \begin{bmatrix} 1 &0  &0 \\0 &0 &0 \\0 &0 &-1 \end{bmatrix}.
\label{spin1}
\end{equation}
The corresponding coarse-graining map is
\begin{equation}
\Phi(A) = \sum_{k\ell} A_{k\ell} j_k j_\ell.
\label{cg:spin1}
\end{equation}

From ~(\ref{esi}) it follows that $\Phi(A)$ is positive semi-definite if and only if  
\begin{equation*}
A=A^* \text{ and } A \le \tr(A).
\label{con1}
\end{equation*}

As any state assigns non-negative values to $\Phi(A)$ with $A \le \tr(A)$ and as conversely any functional which takes the value 1 on $\idty$ and is non-negative on $\Phi(A)$ with $A \le \tr(A)$ extends to a state, the reduced state space is also characterized by the condition
\begin{equation*}
A=A^*\text{ and } A \le \tr(A) \text{ implies } \tr(DA) \ge 0.
\end{equation*}
It is easily seen that this condition is equivalent to
\begin{equation*}
D \text{ density matrix and } D \le \frac{1}{2}.
\end{equation*}

To any density matrix $D$ with $D \le \frac{1}{2}$ we may associate a density matrix 
\begin{equation}
\tilde D = \idty - 2 D.
\end{equation}
The map $D \mapsto \tilde D$ is affine and one to one from the reduced state space to the full state space of $\c M_3$. In particular, every extreme point of the reduced state space corresponds to a pure state. So, every extreme element of the reduced state space is of the form
\begin{equation}
\frac{1}{2}\, (\idty - P),
\end{equation}
where $P$ is a one-dimensional projector in $\c M_3$.

\subsubsection{Infinite spin}

The limiting operational partition when the total angular momentum tends to infinity is given by the relations
\begin{equation}
j_i^* = j_i, \enskip  \bigl[ j_1 \,,\, j_2 \bigr] = 0 \text{ and cycl.\ perm.},\enskip \text{and } j_1^2 + j_2^2 + j_3^2 = 1.
\end{equation}
The Abelian C*-algebra generated by the $j_i$ is just the algebra of continuous complex-valued functions on the unit sphere in $\Rl^3$. An explicit isomorphism is given by
\begin{equation}
j_1(\Omega) = \sin\theta \cos\varphi,\enskip j_2(\Omega) = \sin\theta \sin\varphi,\enskip \text{and } j_3(\Omega) = \cos\theta.
\label{pol}
\end{equation}
Here, $\Omega = (\theta,\varphi)$ is the usual parametrization of a point on the unit sphere in $\Rl^3$ by (co)latitude and longitude: $0 \le \theta \le \pi$ and $0 \le \varphi < 2\pi$. 
Again we introduce the unity preserving map
\begin{equation}
\Phi(A) := \sum_{k\ell} A_{k\ell} j_k j_\ell,
\end{equation}
where $A$ is a complex $3\times3$ matrix. Because of the commutation relations, $\Phi$ is no longer injective. $\Phi(A)$ is positive if and only if 
\begin{equation*}
\Omega \mapsto \sum_{k\ell} A_{k\ell} j_k(\Omega) j_\ell(\Omega) 
\end{equation*}
is a positive function. This is equivalent to impose that
\begin{equation}
\< \bm x \,,\, A\, \bm x \> \ge 0,\enskip \text{for all } \bm x \in \Rl^3 
\label{inf1}
\end{equation}
or, equivalently that $A + A^{\s T} \ge 0$. Writing 
\begin{equation*}
A = \frac{1}{2}\, \bigl( A + A^{\s T} \bigr) + \frac{1}{2} \bigl(A - A^{\s T} \bigr)
\end{equation*}
we see that
\begin{equation*}
\tr DA = \frac{1}{2}\, \tr D\bigl( A + A^{\s T} \bigr) + \frac{1}{2}\, \tr D \bigl(A - A^{\s T} \bigr).
\end{equation*}
As this expression must be non-negative whenever $A + A^{\s T} \ge 0$ we must have that $D = D^{\s T}$. Hence the reduced state space now consists of all 3 dimensional density matrices with real entries. The extreme points are the pure states on $\c M_3$ generated by normalized vectors with real entries. 

We can for this example describe the maximal entropy embedding $\Psi$. Let us denote by $\Omega = (\theta, \phi)$ the usual spherical coordinates on the unit sphere $\c S_2$ in $\Rl^3$ and by $d\Omega$ the normalized invariant surface measure $\frac{1}{4\pi}\, d\phi \sin\theta d\theta$. According to general principles, the maximal entropy embedding $\Psi$ is given by
\begin{equation}
\Psi(D) = \int_{\c S^2} \!d\Omega\, \r e^{\< \bm j \,,\, \Delta\, \bm j\>}\, |\bm j\>\<\bm j|
\label{gaus}
\end{equation}
where $\bm j = (j_1,j_2,j_3)$ and where $\Delta$ is a real symmetric matrix of dimension 3. An explicit formula for $\Delta$ in terms of $D$ is not available but the problem can be simplified. Let $R$ be an orthogonal transformation of $\Rl^3$, then $\bm j \mapsto R\, \bm j$ can be realized by a change of variables $\Omega \mapsto \Omega'$ that preserves the uniform measure. This change of integration variables can therefore be used to diagonalize $\Delta$. A simple inspection shows that in this case the matrix in the right hand side of~(\ref{gaus}) is also diagonal. Therefore, up to an orthogonal transformation, we must only solve~(\ref{gaus}) for diagonal $D$ and $\Delta$.    

\subsection{Quantum description of a classical system}

The general construction  can be repeated for the case when $\c A$ is a commutative algebra and therefore isomorphic to algebra of continuous complex functions on a phase space $\s\Omega$. The state space $S(\Omega)$ of the system is now the space of probability measures on $\Omega$. Hence we have a one to one correspondence between a positive, normalized functional $\omega$ and the probability measure $\mu(dx)$
\begin{equation}
\omega(f) = \int_{\s\Omega} \!\mu(dx)\, f(x),\enskip f \in \c C(\s\Omega).
\label{riesz}
\end{equation}
The partition $\s V$ consists of complex linearly independent functions on $\s\Omega$
\begin{equation}
\s V = \bigl( v_1, v_2, \ldots, v_d \bigr),\enskip v_j \in \c C(\s\Omega).
\label{ex:5:1}
\end{equation}
For a probability measure $\mu$ on $\s\Omega$ with associated functional $\omega$ we obtain the correlation matrix with the elements
\begin{equation*}
D_{ij}= \omega(v^*_j v_i) = \int_{\s\Omega} \!\mu(\r dx)\, \overline{v_j(x)}\, v_i(x).
\end{equation*}  
Introducing the standard basis $\{e_j\}$ in $\Cx^d$ we can write
\begin{equation*}
\bm v(x) := \sum_{j=1}^d v_j(x)\, e_j,\enskip x \in \s\Omega
\end{equation*}
and conclude that
\begin{equation}
D^{\omega} =  \int_{\s\Omega} \!\mu(\r dx)\, \bigl|\bm v(x) \bigr\> \bigl\< \bm v(x) \bigr|.
\label{ex:5:2}
\end{equation}
Therefore the extreme boundary of the reduced state space consists of a closed subset of the rank one positive matrices in $\c M_d$, pure states in the case of a partition of unity. Moreover, any closed convex subset with such a boundary can be realized by suitable limits of choices of partitions as in~(\ref{ex:5:1}).

\section{Composed  generalized subsystems}
\label{s5}

The formalism for generalized subsystems that was presented here extends naturally to composite systems, they are described by higher rank correlation matrices. This has already been used in the context of quantum dynamical entropy and quantum symbolic dynamics~$\cite{af}$. If two systems are described by the partitions $\{v_{\alpha}\}_{\alpha=1}^n$ and $\{w_{k}\}_{k=1}^m$, then the composed system is described by the partition $\{v_{\alpha}w_{k}\}_{\alpha,k}^{mn}$. The elements of a correlation matrix of the composed system are given by
\begin{equation}
D_{\alpha k;\beta\ell} = \omega \bigl( w^*_k v^*_\alpha v_\beta w_\ell \bigr).
\label{quant:2red}
\end{equation}  

\subsection{Generalized entanglement}

When talking about composed systems questions about entanglement naturally arise. Both the notions of ``generalized entanglement'' and ``generalized subsystems'' have recently emerged in the literature~$\cite{{v3},{v1},{v2}}$. The proposed schemes also deal with projections of states of a large system on a low dimensional spaces but the mathematical structures that have been considered are not so rich as these presented here. E.g.\ neither the order structure nor the idea of composition is natural in those schemes.

The basic notion in the conventional approach to entanglement is that of separability. We say that the $n$-party  correlation matrix $D$ is separable if it can be represented as a classical correlation matrix of the form
\begin{equation}
D_{\alpha_1, \ldots, \alpha_n, \beta_1, \ldots, \beta_n} = \int_\Omega \!\mu(dx)\, \overline{f_{\alpha_1}(x)} f_{\beta_1}(x) \cdots \overline{f_{\alpha_n}(x)} f_{\beta_n}(x),
\label{clas-n}
\end{equation}
where $\mu$ is a probability measure on $\Omega$ and $f_\alpha$ are measurable functions. Using the representation~(\ref{ex:5:2}) it is not difficult to show that the above definition is, up to normalization, equivalent to the standard one
\begin{equation}
D = \sum_{k_1, k_2, \ldots, k_n} \lambda\bigl( k_1, k_2, \ldots, k_n \bigr)\, P_{k_1} \otimes P_{k_2} \otimes \cdots \otimes P_{k_n}, 
\end{equation}
where $\lambda\bigl(k_1, k_2, \ldots, k_n \bigr) > 0$ and where the $P_k$'s are one dimensional projectors.

One should notice that for a single system, i.e.\ for $D_{\alpha\beta}$ there always exists a classical representation~(\ref{clas-n}). Indeed as $\Omega$ one can take the manifold of all normalized vectors $\psi$ in the Hilbert $\Cx^d$ and put $f_\alpha(\psi) = \< e_\alpha \,,\, \psi \>$ where $\{e_\alpha\}$ is a basis in $\Cx^d$.

\subsubsection*{An entangled bipartite correlation matrix}

Consider bosonic  systems composed of two parts with corresponding sets of annihilation and creation operators $\bigl( a_k, a^\dagger_k \bigr)$ and $\bigl( b_\alpha , b^\dagger_\alpha \bigr)$, $k, \alpha = 1,2$. The correlation matrix has the form
\begin{equation}
D_{k\alpha,\ell\beta} = \tr \bigl( \omega\, b^\dagger_\ell a^\dagger_\beta a_\alpha b_k \bigr).
\label{ex}
\end{equation}
Consider a two-boson state
\begin{equation*}
|\psi\> = \sum_{k\ell} \gamma_{k\ell} b^\dagger_k a^\dagger_\ell |0\>\otimes|0\>,
\end{equation*}
with the normalization condition
\begin{equation*}
\<\psi \,,\, \psi\> = \sum_{k\ell} \abs{\gamma_{k\ell}}^2 = \tr G G^\dagger = 1,
\end{equation*}
where  $G$ denotes the  $2\times 2$ matrix $[\gamma_{k\ell}]$ .
For such a state the correlation matrix (\ref{ex}) has the form 
\begin{equation}
\Omega =
\begin{bmatrix}
\abs{\gamma_{11}}^2 & \gamma_{11} \overline{\gamma_{12}} & \gamma_{11} \overline{\gamma_{21}} & \gamma_{11} \overline{\gamma_{22}} \\
\gamma_{12} \overline{\gamma_{11}} & \abs{\gamma_{12}}^2 & \gamma_{12} \overline{\gamma_{21}} & \gamma_{12} \overline{\gamma_{22}} \\
\gamma_{21} \overline{\gamma_{11}} & \gamma_{21} \overline{\gamma_{12}} & \abs{\gamma_{21}}^2 & \gamma_{21} \overline{\gamma_{22}} \\
\gamma_{22} \overline{\gamma_{11}} & \gamma_{22} \overline{\gamma_{12}} & \gamma_{22} \overline{\gamma_{21}} & \abs{\gamma_{22}}^2
\end{bmatrix}.
\label{ex1}
\end{equation}
and partial transposition of~(\ref{ex1}) yields 
\begin{equation}
\Omega^{\idty\otimes\textsf T} =
\begin{bmatrix}
\abs{\gamma_{11}}^2 & \gamma_{12} \overline{\gamma_{11}} & \gamma_{11} \overline{\gamma_{21}} & \gamma_{12} \overline{\gamma_{21}} \\
\gamma_{11} \overline{\gamma_{12}} & \abs{\gamma_{12}}^2 & \gamma_{11} \overline{\gamma_{22}} & \gamma_{12} \overline{\gamma_{22}} \\
\gamma_{21} \overline{\gamma_{11}} & \gamma_{22} \overline{\gamma_{11}} & \abs{\gamma_{21}}^2 & \gamma_{22} \overline{\gamma_{21}} \\
\gamma_{21} \overline{\gamma_{12}} & \gamma_{22} \overline{\gamma_{12}} & \gamma_{21} \overline{\gamma_{22}} & \abs{\gamma_{22}}^2
\end{bmatrix}.
\label{ex1t}
\end{equation}
The eigenvalues of $\Omega^{\idty\otimes\textsf T}$ are given by
\begin{equation*}
\pm\det( \abs G)\enskip \text , \enskip \frac{1}{2}\bigl( 1 \pm \sqrt{1- 4(\det(\abs G))^2}\bigr)
\end{equation*}
where $\abs G = \sqrt{GG^\dagger}$. So, for $\det(G) \neq 0$, one eigenvalue of $\Omega^{\idty\otimes\textsf T}$ is negative and, according to the criterion of partial transposition~\cite{ah,ap}, this means that the correlation matrix $\Omega$ is entangled.
    
\subsection{Entanglement in time}

This idea appeared for the first time in the paper by Leggett and Garg~\cite{lg} who proposed to check Bell's inequality for correlations  corresponding to projective measurements at different times. The problem of a ``quantumness'' test for temporal correlations can be easily formulated in our language of correlation matrices. If the evolution of the system from time $0$ to $t$ is described in Heisenberg picture by the completely positive unity preserving map $\Lambda_t$, then we can define a time-dependent correlation function, normalized for partitions of unity, by
\begin{equation}
D_{k\ell, k'\ell'}(t) = \tr \bigl( \omega\, v^*_k \Lambda_t(v^*_\ell v_{\ell'}) v_{k'} \bigr).
\label{extime}
\end{equation}
One can now apply well-known criteria of separability or measures of entanglement to describe the evolution of ``quantumness'' encoded in the correlation matrices of the single system. Examples will be discussed in a future publication. One should notice that this approach is related to the formalism of thermal Green functions in statistical mechanics or quantum field theory~\cite{AGD}. In those cases the reference state $\omega$ is either thermal or the vacuum.

\section{Conclusions}

In this paper we introduced and examined generalized subsystems (GS) as a unifying formalism for the reduced description of complex and open quantum systems. It covers a number of known examples, like the standard approach to open systems with tensor product structure, single particle descriptions of many-body systems and Green functions methods. The new examples of GS's are quantum systems with symmetries described by Lie algebras. Our approach fits well with a large class of approximate evolution equations and with the state estimation based on the maximal entropy principle. The mathematical formalism involves completely positive maps and correlation functions, it has a rich mathematical structure including order relations and compositions. This yields a natural notion of generalized entanglement in space and time, this issue has only slightly been touched and will be investigated in the future. 

\noindent
\textbf{Acknowledgements}
This work is partially funded by the Belgian Interuniversity Attraction Poles Programme P6/02 (MF). Support by the Polish-Flemish bilateral grant BIL~05/11 (MP) and by the Polish research network LFPPI (RA, MP) is also acknowledged .

\end{document}